\def\lsi{\raise0.3ex\hbox{$<$\kern-0.75em\raise-1.1ex\hbox{$\sim$}}}
\def\gsi{\raise0.3ex\hbox{$>$\kern-0.75em\raise-1.1ex\hbox{$\sim$}}}

\newcommand{\gsim}{\mathop{\gsi}}

\documentstyle[twoside,fleqn,espcrc2,epsf]{article}
\title{Electroweak Phase Transitions}
\author{Z. Fodor \address{Institute for Theoretical Physics, E\"otv\"os
University,\\ H-1088 Budapest, HUNGARY}}
\begin{document}
\begin{abstract}
Recent developments on the four dimensional (4d) lattice studies of the
finite temperature electroweak phase transition (EWPT) are summarized. 
The phase diagram is given in the continuum limit. The finite temperature
SU(2)-Higgs phase transition is of first order for Higgs-boson masses
$m_H<66.5 \pm 1.4$ GeV.   
Above this endpoint only a
rapid cross-over can be seen. The full 4d result
agrees completely with that of the dimensional reduction
approximation. The Higgs-boson endpoint
mass in the Standard Model (SM) would be  $72.1 \pm 1.4$ GeV. Taking into account
the LEP Higgs-boson mass lower bound excludes
any EWPT in the SM.
A one-loop calculation of the static potential in the SU(2)-Higgs model
enables a precise comparison between lattice
simulations and perturbative results. The most popular extension
of the SM, the Minimal Supersymmetric SM (MSSM) is 
also studied on 4d lattices.
\end{abstract}
\maketitle

\section{INTRODUCTION}

The visible Universe is made of matter. This fact is based on
observations of the cosmic diffuse $\gamma$-ray background, which
could be larger than the present limits, if boundaries between
``worlds'' and ``antiworlds'' had existed \cite{cohen98}.
The observed baryon asymmetry of the universe was eventually determined
at the EWPT \cite{KRS85}. On the one
hand this phase transition was the last instance during which baryon 
asymmetry could have been generated around $T \approx 100$ GeV, 
on the other hand at these temperatures any B+L asymmetry 
could have been washed out. The possibility of baryogenesis at
the EWPT is a particularly attractive 
one, since the underlying physics can be --and already largely has been-- 
tested at collider experiments. Thus, the detailed understanding of 
this phase transition is very important. 

A succesfull baryogenesis scenario consists three ingredients,
the Sakharov's conditions. \\
1. Baryon number violating processes\\
2. C and CP violation\\
3. Departure from equilibrium.\\
All of the three conditions has non-perturbative features and are
studied on the lattice (e.g. at this conference baryon number violating
sphalerons have been discussed by \cite{moore99}, spontaneous CP 
violation by \cite{laine99}, whereas this contribution  mostly 
studies the out of equilibrium condition).

It is rather easy to see the necessity of the first two
conditions. Without baryon number violation no net baryon asymmetry can be
generated. C and CP violation are needed to give a direction to
the processes. The standard picture concerning the third
condition is the turn-off of the baryon number violating rate
after the phase transition, which means a smaller sphaleron rate than
the Hubble rate. Inspecting the formula for the sphaleron rate
one needs a strong enough phase transition, thus $v/T_c \gsim 1$. This 
ratio is of particular interest and both perturbative
and lattice studies have the main goal to determine it.

The first-order nature of the EWPT for light Higgs
bosons can be shown within perturbation theory.  However,
perturbation theory breaks down for Higgs boson masses ($m_H$) larger than about
60 GeV due to bad infrared behavior of the gauge-Higgs part of
the electroweak theory \cite{perturb}. Solutions of gap-equations
even suggest an end-point scenario for the first order
EWPT \cite{gap}.  Numerical simulations
are needed to analyze the nature of the transition
for realistic Higgs bosons.

One very succesfull possibility is to construct an effective three 
dimensional (3d)  theory by using
dimensional reduction, which is a perturbative step. The non-perturbative study
is carried out in this effective 3d model \cite{3d-sim}.
The end-point of the phase transition is determined and 
its universality class is studied \cite{3d-end}.

Another approach is to use 4d simulations. The complete lattice
analysis of the SM is not feasible due to the presence of chiral
fermions, however, the infrared problems are connected only with the bosonic
sector. These are the reasons why the problem is usually studied by simulating
the SU(2)-Higgs model on 4d lattices, and perturbative steps are
used to include the U(1) gauge group and the fermions. Finite temperature
simulations are carried out on lattices with volumes $L_t \cdot L_s^3$, where 
$L_t \ll L_s$ are the temporal and spatial extensions of the lattice, 
respectively. Systematic studies were carried out 
for $m_H \approx$ 20 GeV, 35 GeV, 50 GeV and 75 GeV
\cite{4d}. The lattice spacing is basically fixed by the number of the 
lattice points in the temporal direction
($T_c=1/(L_t a)$, where $T_c$ is the critical temperature in physical units);
therefore huge lattices are needed to study the soft modes. This
problem is particularly severe for Higgs boson masses around the W mass,
for which the phase transition is weak and typical correlation lengths
are much larger than the lattice spacing. In this case asymmetric lattice
spacings are used \cite{4d-asym}.

\section{END-POINT IN FOUR DIMENSIONS}

The 4-d SU(2)-Higgs model is studied 
on both symmetric  and asymmetric \cite{4d-asym}
lattices, i.e. lattices
with equal or different spacings in temporal ($a_t$) and spatial
($a_s$) directions. 
The asymmetry of the
lattice spacings is given by the asymmetry factor $\xi=a_s/a_t$.
The different lattice spacings can be ensured by
different coupling strengths in the action for time-like and space-like
directions. The action reads in standard notation \cite{4d}
\begin{eqnarray}
&& S[U,\varphi]= \nonumber \\
&& \beta_s \sum_{sp}
\left( 1 - {1 \over 2} {\rm Tr\,} U_{pl} \right)
+\beta_t \sum_{tp}
\left( 1 - {1 \over 2} {\rm Tr\,} U_{pl} \right)
\nonumber \\
&&+ \sum_x \left\{ {1 \over 2}{\rm Tr\,}(\varphi_x^+\varphi_x)+
\lambda \left[ {1 \over 2}{\rm Tr\,}(\varphi_x^+\varphi_x) - 1 \right]^2
\right. \nonumber \\
&&\left.
-\kappa_s\sum_{\mu=1}^3
{\rm Tr\,}(\varphi^+_{x+\hat{\mu}}U_{x,\mu}\,\varphi_x) \right. \nonumber \\
&& \left. -\kappa_t {\rm Tr\,}(\varphi^+_{x+\hat{4}}U_{x,4}\,\varphi_x)\right\},
\end{eqnarray}
We introduce
$\kappa^2=\kappa_s\kappa_t$ and
$\beta^2=\beta_s\beta_t$. 
The anisotropies
$\gamma_\beta^2=\beta_t/\beta_s$ and $\gamma_\kappa^2=\kappa_t/\kappa_s$
are functions of $\xi$. We use
$\xi=4.052$, which corresponds to $\gamma_\kappa=4$ and $\gamma_\beta=3.919$.

\begin{figure}
\centerline{\epsfxsize=0.9 \linewidth \epsfbox{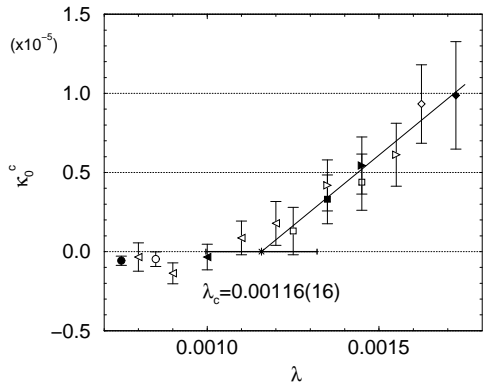}}
\vspace{-0.9cm}
\caption{Imaginary part of first Lee-Yang zero
as a function of $\lambda$
from simulations on symmetric lattices with $L_t=2$.
Filled symbols are without $\lambda$-reweighting,
while open symbols with
$\lambda$-reweighting from filled symbol with same shape.} 
\vspace{-0.5cm}
\end{figure}

The determination of the end-point of the finite temperature
EWPT is done by the use of the Lee-Yang zeros of the
partition function ${\cal Z}$.
Near the first order phase transition point the partition function reads
${\cal Z}={\cal Z}_s + {\cal Z}_b \propto \exp (-V f_s)+ \exp (-V f_b) \ ,$
where the indices s(b) refer to the symmetric (broken) phase and $f$ stands
for the free-energy densities. We also have
$f_b = f_s + \alpha (\kappa - \kappa _c ) \ ,$ ,
since the free-energy density is continuous. It follows that
${\cal Z} \propto  \exp [ -V ( f_s +f_b )/2 ] 
\cosh [ -V \alpha (\kappa -\kappa_c )] \ ,$
which shows that for complex $\kappa$ ${\cal Z}$ vanishes at
${\rm Im} (\kappa )=2 \pi \cdot (n-1/2) / (V\alpha )$
for integer $n$.  In case a first order phase transition is present,
these Lee-Yang
zeros move to the real axis as the volume goes to infinity. In case a
phase transition is absent the Lee-Yang
zeros stay away from the real $\kappa $ axis. 
Denoting
$\kappa_0$ the lowest zero of ${\cal Z}$, i.e. the  position of the zero
closest to the real axis, one expects in the vicinity of
the end-point the scaling law
${\rm Im}(\kappa_0)=C(L_t,\lambda)V^{-\nu}+\kappa_0^c(L_t,\lambda)$.
In order to pin down the end-point we are looking
for a $\lambda$ value for which $\kappa_0^c$ vanishes.
In practice we analytically continue ${\cal Z}$ to complex
values of $\kappa $ by reweighting.
Small changes  in
$\lambda$ were taken into account by reweighting.
The dependence of $\kappa_0^c$ on  $\lambda$ 
\cite{Aoki99} is shown in fig. 1.
To determine the critical value of $\lambda$
i.e. the largest value, where $\kappa_0^c=0$, we
have performed  fits linear in $\lambda$ to the non-negative
$\kappa_0^c$ values.
\begin{figure}[t]
\centerline{\epsfxsize=0.85 \linewidth \epsfbox{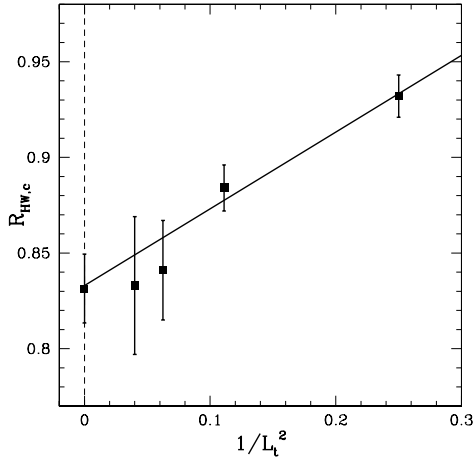}}
\vspace{-0.7cm}
\caption{Dependence of $R_{HW,c}$, on $1/L_t^2$ and extrapolation to the
continuum limit.}
\vspace{-0.5cm}
\end{figure}

In the isotropic case \cite{Aoki99}, we have used $L_t =2$.
The Lee-Yang analysis gave $\lambda_c =0.00116(16)$ for the end-point.
Performing $T=0$ simulations with the same parameters this can be converted
to $m_{H,c}=73.3 \pm 6.4$GeV.
In the anisotropic lattice simulation case \cite{Csikor99}
we also performed a continuum extrapolation for $L_t=2,3,4,5$ (fig. 2),
moving along the lines of constant physics (LCP), and obtained
$66.5 \pm 1.4$ GeV, which is our final result for the end-point in the
SU(2)-Higgs model.

Based on previous 4d simulation results one can determine the 
phase diagram of the finite temperature EWPT
and compare it with the 3d analysis (fig. 3.)
as it has been done in ref. \cite{Laine99}. The phase transition lines 
$T_c(m_H)$,
are in perfect agreement for $m_H\gsim 25$ GeV.
For strong first order phase transition close to the
Coleman-Weinberg limit the 3d approach seems to be less
accurate.  The error bars on the endpoints are on the
few percent level, thus uncertainty of the dimensional reduction
around the end-point is also in this range. 
This indicates that the analogous perturbative
inclusion  of the fermions results also in few percent error on
$m_H$.

\begin{figure}[t]
\centerline{\epsfxsize=1.10 \linewidth \epsfbox{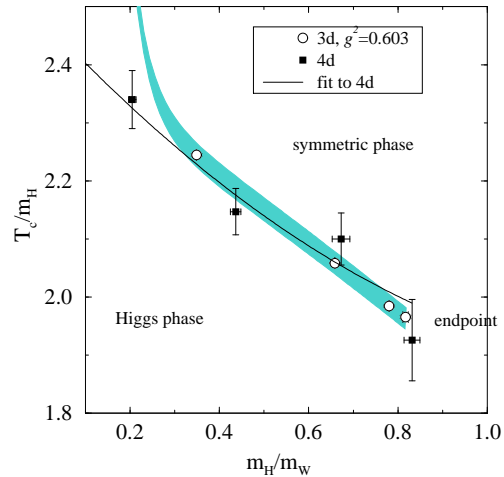}}
\vspace{-0.7cm}
\caption{
A comparison of the phase diagrams
obtained from direct 4d simulations (squares) and from dimensionally 
reduced 3d  simulations (shaded region).}
\vspace{-0.5cm}
\end{figure}

One can determine what is the endpoint value in the full SM.
As it was shown previously the perturbative integration of
the heavy modes is correct within our error bars. Therefore we use
perturbation theory to transform the SU(2)-Higgs model endpoint
value to the full SM. We obtain $72.1 \pm 1.4$ GeV, where the 
dominant error
comes from the measured error of $R_{HW,cont.}$. The error
on $g_R^2$ is eliminated by calculating the relationship
between the coupling definitions used in perturbation theory
(${\overline {\rm {MS}}}$) and lattice simulations
(from static potential) \cite{Laine99,Csikor99a}. The calculation
of this relationship and a comparison of the perturbative and lattice
results on the EWPT will be shortly discussed
in the next section.

The full SM result needs some explanation. Based on vacuum stability the
measured top mass ($m_{top} \approx 175$ GeV) results in a lower bound for
the Higgs boson mass (approx. 130 GeV). This value is higher than the 
previously mentioned $72.1 \pm 1.4$ GeV. For the pure SU(2)-Higgs
model the endpoint Higgs mass is $66.5 \pm 1.4$ GeV. The inclusion of the
fermions, especially the top increases the endpoint slightly. For a 
hypothetical top quark mass less than 
approximately 150 GeV the lower bound is less than $\approx 70$ GeV, thus
it is below the endpoint and it gives a reliable theory. Increasing the
top quark mass the lower bound gets larger than the endpoint. This means
that independently of the direct experimental bounds on the
Higgs boson mass no EWPT  exists in the
SM. 

\section{RELATIONSHIP BETWEEN GAUGE COUPLINGS}

Despite the fact that the perturbative and lattice approaches
are systematic and well-defined, it is not easy to compare
their predictions. The reason is that in lattice simulations
the gauge coupling constant is determined from the static potential, whereas
in perturbation theory the ${\overline {\rm {MS}}}$ scheme is used. One 
can calculate the static potential on the one-loop level in the
SU(2)-Higgs model \cite{Laine99,Csikor99a}. As expected the numerical
difference between the
two conventions is not that large, it is within a few percent, for details
see  \cite{Laine99,Csikor99a}.
With this connection we could perform a precise comparison
between the predictions of perturbative and lattice approaches (fig. 4).

In \cite{Csikor99a}  the existing lattice data was reanalyzed and a
continuum limit extrapolation was performed whenever it was possible. 
The only quantity which is measured so precisely that the definition
of the gauge coupling constant is essential is the ratio of the
critical temperature to the Higgs boson mass. As it has been
observed already for $M_H \approx 35$ GeV the perturbative
value of $T_c$ is larger than in lattice simulations.
This sort of discrepancy disappears for larger
Higgs boson masses. A plausible reason for this fact
is the convergence of the high temperature expansion
used in the perturbative approach.

The most dramatic differences appear clearly as we get closer to the end
point. The perturbative approach gives non-vanishing jump of the
order parameter, non-vanishing latent heat and interface tension, while
the lattice results suggest rapid decrease of these quantities
and no phase transition beyond the end-point.

\begin{figure}[t]
\centerline{\epsfxsize=1.15 \linewidth \epsfbox{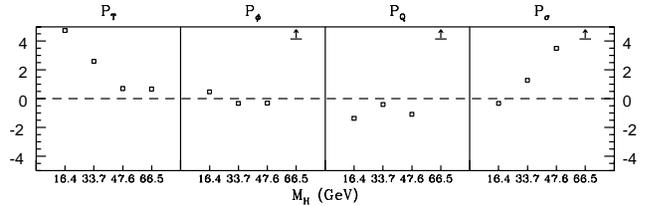}}
\vspace{-0.7cm}
\caption{Pulls plotted against the Higgs mass for $T_c$, $v/T$, latent heat
and interface tension.}
\vspace{-0.5cm}
\end{figure}

\section{PHASE TRANSITION IN THE MSSM}

As it was demonstrated in the previous sections the SM
is not suitable for baryogenesis, not even for a first order
EWPT. Several extended models were studied
in order to obtain a stronger first order phase transition and
a reliable baryon asymmetry. The most popular model
is the MSSM, which perturbatively shows a much stronger
phase transition than the SM \cite{mssm} (even an intermediate colour
breaking phase transition is possible in these scenarios)
. Lattice studies in a 3d reduced model (with one 
Higgs doublet) basically confirmed the perturbative results
\cite{mssm3d}.

We performed a 4d lattice study with the bosonic sector of the
MSSM \cite{mssm4d}. The lattice
action is too long to be presented here, thus only
the fields involved are listed. Both of the Higgs doublets,
the stop, sbottom scalars and  SU(2), SU(3)
gauge fields were included. It is of particular importance to 
keep both of the Higgs doublets, since according to the standard
scenario the generated 
baryon asymmetry is directly proportional to the change of the 
ratio of their expectation values
$n_B \propto \langle v^2 \rangle \Delta \beta(T_c)$.
Here the length squared of the Higgs field ($v^2=v_1^2+v_2^2$)
is integrated over the bubble wall. 
The ratio of the expectation values of the two Higgs fields is 
$\tan \beta=v_1/v_2$, and the difference between the $\beta$ 
values are taken in the ``symmetric'' and in the ``broken''
phases. 

We had simulations at $L_t=2,3,4,5$ and moved along the line of constant
physics. Our simulation point corresponds to $\tan\beta(T=0) \approx 6$, 
and the mass of the lightest Higgs bosons is approx. 35 GeV
(in the bosonic theory). 

Two values of $\alpha_s$ were taken
(the physical and a smaller one). The physical $\alpha_s$
resulted in $v/T_c \approx 1.5$, whereas the smaller
value of $\alpha_s$ gave a stronger phase transition
$v/T_c \approx 2$. Perturbation theory predicts just the
opposite behaviour (stop-gluon setting sun graphs are proportional
to the strong coupling and they are responsible for the strengthening of the
phase transition). The reason can be the difference between the 
renormalization effects in the stop sector. 

We measured the $\beta$ parameter in both phases at the phase transition.
One obtains $\tan^2 \beta(sym) = 38.13(15)$, $\tan^2
\beta(brok) = 36.04(15)$, which gives $\Delta \beta = 0.0045(7)$. This result
is far below the perturbative prediction $\Delta \beta (pert.)=0.017$.

\section{CONCLUSIONS}

The endpoint of hot EWPT with the technique of Lee-Yang zeros 
from simulations in 4d SU(2)-Higgs model was determined.
The phase transition is first order for Higgs masses less than
$66.5 \pm 1.4$ GeV, while for larger Higgs masses only a rapid cross-over
is expected. The phase diagram of the model was given.

It was shown non-perturbatively that for the bosonic sector of the 
SM the dimensional reduction procedure works within a few 
percent. This indicates that the analogous perturbative
inclusion of the fermionic sector results also in few percent error.
In the full SM we get $72.1 \pm 1.4$ GeV for the end-point, 
which is below the lower experimental bound. This fact is a clear
sign for physics beyond the SM.

Based on a one-loop calculation on the static potential of the
SU(2)-Higgs model a direct comparison between the perturbative
and lattice results was performed. 

The MSSM is more promising for a succesfull baryogenesis. Some
4d results were shown, indicating a strong first order
phase transition. 

{\bf Acknowledgments:} This work was partially supported by
Hung. Grants No. 
OTKA-T22929-29803-M28413-FKFP-0128/1997.


\begin{thebibliography}{10}
\bibitem{cohen98} A.G. Cohen, A. De Rujula, S.L. Glashow,
Astrophys. J. 495 (1998) 539. 
\bibitem{KRS85} V.A. Kuzmin, V.A. Rubakov and M.E. Shaposhnikov,
Phys. Lett. B155 (1985) 36.
\bibitem{moore99} D. B\"odeker, G.D. Moore, K. Rummukainen, hep-lat/9909054
(these proceedings).
\bibitem{laine99} M. Laine and K. Rummukainen, hep-lat/9908045 (these
proceedings).
\bibitem{perturb} P. Arnold and O. Espinosa, Phys. Rev. D47 (1993) 3546, 
Erratum ibid. D50 (1994) 6662;
Z. Fodor and A. Hebecker, Nucl. Phys. B432 (1994) 127;
W. Buchm\"uller, Z. Fodor, and A. Hebecker, Nucl. Phys. B447 (1995) 317.
\bibitem{gap}  W. Buchm\"uller et al., Ann.\ Phys.\ (NY) 234 (1994) 260;
W. Buchm\"uller, O. Philipsen, Nucl. Phys. B443 (1995) 47.
\bibitem{3d-sim} K. Farakos et al., Nucl. Phys.  B425 (1994) 67;
A. Jakov\'ac, K. Kajantie and A. Patk\'os, Phys.  Rev. D49 (1994) 6810;
K. Kajantie et al., Nucl. Phys. B458 (1996) 90; ibid. B466 (1996) 189; 
Phys. Rev. Lett. 77 (1996) 2887.
\bibitem{3d-end} F. Karsch et al., Nucl. Phys. Proc. Suppl. 53 (1997) 623;
M. G\"urtler et al., Phys. Rev. D56 (1997) 3888 (1997);
K. Rummukainen et al., Nucl. Phys. B532 (1998) 283.
\bibitem{4d}  Z. Fodor et al., Phys. Lett. B334 (1994) 405; Nucl. Phys. 
B439 (1995) 147; F. Csikor et al., Nucl. Phys. B474 (1996) 421;
Phys. Lett. B357 (1995) 156.
\bibitem{4d-asym} F. Csikor, Z. Fodor, Phys. Lett. B380 (1996) 113;
F. Csikor, Z. Fodor and J. Heitger, Phys. Rev. D58 (1998) 094504.
\bibitem{Aoki99} Y. Aoki et al., Phys. Rev. D60 (1999) 013001.
\bibitem{Csikor99} F. Csikor, Z. Fodor and J. Heitger, 
Phys. Rev. Lett. 82 (1999) 21.
\bibitem{Laine99} M. Laine,  JHEP 9906 (1999) 020, hep-ph/9903513.
\bibitem{Csikor99a} F. Csikor et al., hep-ph/9906260.  
\bibitem{mssm} A. Brignole et al., Phys. Lett. B324 (1994) 181;
M. Carena, M. Quiros, C.E.M. Wagner,  Phys. Lett. B380 (1996) 81;
B. de Carlos, J.R. Espinosa, Nucl. Phys. B503 (1997) 24;
D. B\"odeker et al., Nucl. Phys. B497 (1997) 387;
M. Losada, Nucl. Phys. B537 (1999) 3. 
\bibitem{mssm3d} M. Laine, K. Rummukainen, Phys. Rev. Lett. 80
(1998) 5259; Nucl. Phys. B535 (1998) 423.   
\bibitem{mssm4d} F. Csikor, Z. Fodor, P. Heged\H us, V. Horv\'ath,
S.D. Katz, A. Pir\'oth, in preparation.
\end{thebibliography}
\end{document}